\documentclass[useAMS,usenatbib]{mn2e}
\usepackage{epsfig}
\usepackage{color} 
\usepackage{natbib}
\usepackage[colorlinks=true,citecolor=blue]{hyperref}
\title[SED of PKS 2155-304]{Spectral Energy Distributions of the BL Lac PKS 2155$-$304 from XMM-Newton}

\author[Bhagwan et al.]
{Jai Bhagwan$^{1,2}$,
Alok C.\ Gupta$^{1}$\thanks{E-mail: acgupta30@gmail.com}
I. E. Papadakis$^{3, 4}$, 
Paul J. Wiita$^{5}$
\\
$^{1}$Aryabhatta Research Institute of Observational Sciences (ARIES),
Manora Peak, Nainital -- 263002, India\\
$^{2}$School of Studies in Physics \& Astrophysics, Pt.\ Ravishankar Shukla University, Amanaka G.E. Road, Raipur -- 492010, India \\
$^{3}$Department of Physics and Institute of Theoretical and Computational Physics, University of Crete, GR-71003 Heraklion, Greece \\
$^{4}$IESL, Foundation for Research and Technology, 71110 Heraklion, Greece \\
$^{5}$Department of Physics, The College of New Jersey, PO Box 7718, Ewing, NJ 08628-0718, USA \\
}

\begin{document}

\date{Accepted 2014 August 15. Received 2014 August 15; in original form 2014 March 25}

\pagerange{\pageref{firstpage}--\pageref{lastpage}} \pubyear{2014}

\maketitle

\label{firstpage}

\begin{abstract}
We have used all 20 archival XMM-Newton observations  of PKS $2155-304$  with {\it simultaneous}  X-ray and UV/optical data to study its long term  flux and spectral variability.  We find significant variations, in all bands, on time scales of years with an rms amplitude of $\sim 35-45$ per cent, though the optical/UV variations are not correlated with those in the X-ray.  We  constructed spectral energy distributions  (SEDs) that span more than three orders of magnitude in frequency and we first fitted them with a log-parabolic model; such models have been applied many times in the past for this, and  other, blazars.  These fits were poor, so we then examined combined power-law and log-parabolic fits that are improvements.  These models  indicate that the optical/UV and X-ray flux variations are mainly driven by model normalization variations, but the X-ray band flux is also affected by spectral variations, as parametrized with the model ``curvature" parameter, $b$.  Overall, the energy at which the emitted power is maximum  
correlates positively with the total flux.   As the spectrum shifts to higher frequencies, the spectral ``curvature" increases, in contrast to what is expected if a single log-parabolic model were an acceptable representation of the broad band SEDs. Our results suggest that  the optical/UV and X-ray emissions in this source may arise from different lepton populations.  
\end{abstract}


\section{Introduction}

Blazars comprise a subclass of radio-loud active galactic nuclei that consist of BL Lacertae objects (BL Lacs) 
and flat spectrum radio quasars (FSRQs). These objects show flux and polarization variability on diverse time 
scales across the entire electromagnetic (EM) spectrum (Ulrich et al 1997). According to the current paradigm, blazars have supermassive 
black holes at their centers that accrete material and produce relativistic jets that happen to be oriented close to our line 
of sight (Urry \& Padovani 1995). The observed spectra of blazars are dominated by nonthermal radiation 
 produced by relativistic electrons spiraling around the  magnetic fields in  relativistic jets 
(Blandford \& Rees 1978; Urry \& Padovani 1995). 

The spectral energy distributions (SEDs) of blazars have two broad humps in the
log$(\nu F_{\nu})$ vs log$(\nu)$ representation (Ghisellini et al.\ 1997). The low energy SED hump peaks 
in a frequency ranging from sub-mm to soft X-ray bands is well explained by synchrotron emission 
from an ultra-relativistic electron population residing in the magnetic fields of the approaching relativistic jet (e.g., Maraschi 
et al.\ 1992; Ghisellini et al.\ 1993; Hovatta et al.\ 2009; and references therein). The high energy 
SED hump that peaks in the MeV--TeV gamma-ray bands,  is usually attributed to  inverse Compton (IC) 
scattering of  photons off those relativistic electrons. 
Depending on the peak frequency of synchrotron hump, $\nu_{s}$, blazars are often further classified 
into three categories: low synchrotron-peaked (LSP) blazars have $\nu_{s} \leq 10^{14}$ Hz, 
intermediate-synchrotron-peaked (ISP) have $10^{14}\leq \nu_{s} \textless 10^{15}$ Hz, and high 
synchrotron-peaked (HSP) blazars have $\nu_{s}\textgreater 10^{15}$ Hz (Abdo et al. 2010). 

In the present work, we have studied the optical/UV and X-ray band variability of BL Lac PKS $2155-304$ which is a HSP. 
PKS $2155-304$ was one of the first 
recognized BL Lacs  (Schwartz et al.\ 1979, Hewitt \& Burbidge 1980) and is the brightest object in 
UV to TeV energies in the southern hemisphere. The redshift of PKS $2155-304$ is $0.116\pm 0.002$ as 
determined by optical spectroscopy of the galaxies in the BL Lac field (Falomo et al.\ 1993). This 
object has been studied on many occasions in single and multiple bands of the EM spectrum to search for
variability, cross-correlated variability, SEDs and other properties of the source on  diverse 
timescales (e.g., Shimmins \& Bolton 1974; Carini \& Miller 1992; Urry et al.\ 1993; Brinkmann et al.\ 
1994; Marshall et al.\ 2001; Aharonian et al.\ 2005b; Dominici et al.\ 2006; Dolcini et al.\ 2007; Piner 
et al.\ 2008; Sakamoto et al.\ 2008; Kastendieck et al.\ 2011; Abramowski et al.\ 2012, and references 
therein). The first X-ray observation of PKS $2155-304$ was taken by Schwartz et al.\ (1979) using HEAO-1. 
Chadwick et al.\ (1999) detected for the first time very high energy gamma-ray photons from 
this source using the Durham MK 6 telescope and thus classified it as a TeV blazar. There also are more recent claims of
TeV emission from this source  (Abramowski et al.\ 2012, and references therein).
Simultaneous multi-band observations of the source from optical to X-ray bands using XMM-Newton data 
were reported by Zhang et al.\ (2006a, 2006b). Zhang (2008) found that the synchrotron emission of 
PKS 2155-304 peaked in the UV-EUV bands rather then the soft X-ray band. Gaur et al.\ (2010)  searched for
intra-day variability and quasi-periodic oscillations (QPOs) in the source using XMM-Newton data. 
Using International Ultraviolet Explorer  observations in the UV band of PKS $2155-304$ Urry et al.\ 
(1993) reported a possible short-lived QPO of $\sim$ 0.7 day.  More recently, stronger evidence for a $\sim$ 4.6 
hr QPO in this source on one occasion in XMM-Newton observations  was reported (Lachowicz et al.\ 2009).

\begin{table*}
\begin{center}
{\bf Table 1.} Observation log of PKS 2155-304 with XMM-Newton EPIC/pn and optical monitor   \\
\begin{tabular}{ccrcccccr}  \hline  
Revolution& Obs.ID &Exp.ID& Start Date& End Date & Duration&Pileup&OM Filters$^{1}$ \\ 
&&&&&(ks)&& \\ \hline
087&0124930101&087-1&2000-05-30 05:29:42&2000-05-30 22:28:11&37.9&Yes&3 \\
087&0124930201&087-2&2000-05-31 00:30:51&2000-05-31 20:40:09&59.3&Yes&1\\
0174&0080940101&0174-1&2000-11-19 18:38:20&2000-11-20 11:26:51&57.2&Yes&1\\
0174&0080940301&0174-2&2000-11-20 12:53:01&2000-11-21 05:56:32&58.1&Yes&1 \\
0362&0124930301&0332-1&2001-11-30 02:36:09&2001-12-01 04:19:46&44.6&Yes&1,2,3,4,5,6 \\
0450&0124930501&0450-1&2002-05-24 09:31:02&2002-05-25 14:38:50&96.1&Yes&3,4,5,6 \\
0545&0124930601&0545-1&2002-11-29 23:27:28&2002-12-01 07:18:43&56.8&No&1,2,3,4,5,6  \\
0724&0158960101&0724-1&2003-11-23 00:46:22&2003-11-23 08:19:01&26.6&No& 2,3,4\\
0908&0158960901&0908-1&2004-11-22 21:35:30&2004-11-23 05:37:29&28.4&No&4,5,6\\
0908&0158961001&0908-2&2004-11-23 19:45:55&2004-11-24 06:59:34&39.9&No&1,2,3,4\\
0993&0158961101&0993-1&2005-05-12 12:51:06 &2005-05-12 20:52:56&26.1&Yes&1,2,3,4,5,6 \\
1095&0158961301&1095-1&2005-11-30 20:34:03&2005-12-01 13:20:58&59.9&Yes &1,2,3,4,5,6 \\
1171&0158961401&1171-1&2006-05-01 12:25:55&2006-05-02 06:26:09&64.3&Yes&1,2,3,4,5,6  \\
1266&0411780101&1266-1&2006-11-07 00:22:47&2006-11-08 04:26:19&29.9&No&1,2,3,4,5,6 \\
1349&0411780201&1349-1&	2007-04-22 04:07:23&2007-04-22 22:59:14&58.5&Yes&1,2,3,4,5,6  \\
1543&0411780301&1543-1&	2008-05-12 15:02:34&2008-05-13 08:02:50&60.7&Yes&1,2,3,4,5,6  \\
1734&0411780401&1734-1&	2009-05-28 08:08:42&2009-05-29 02:09:02&64.3&Yes &1,2,3,4,5,6 \\
1902&0411780501&1902-1&	2010-04-28 23:47:42&2010-04-29 20:26:00&69.1&No &1,2,3,4,5,6 \\
2084&0411780601&2084-1&2011-04-26 13:50:40&2011-04-27 07:34:18&63.3&Yes&1,2,3,4,5,6 \\
2268&0411780701&2268-1&2012-04-28 00:48:26&2012-04-28 19:54:01&53.6&No &1,2,3,4,5,6 \\
\hline

\end{tabular}
\end{center}
\noindent
$^{1}$ 1 = UVW2, 2 = UVM2, 3 = UVW1, 4 = U, 5 = B, 6 = V
\end{table*}

The blazar's flux is rapidly variable in all the EM bands and is often accompanied by spectral changes as well. Changes in the SEDs are very likely associated with changes in the spectra of the emitting electrons.   
Modeling of broad-band SEDs of blazars is required to understand the extreme physical conditions 
inside the different emission regions. Flux variability studies can in principle play an important 
role in understanding the physical phenomena that are responsible for the low, high and outburst states 
of the source. Such studies are very important in discriminating between the models and applying  
tight constraints on model parameters, which are usually changed under the assumption that all other parameters
are fixed (e.g., Mukherjee et al.\ 1999; Petry et al.\ 2000; Hartman et al.\ 2001). In the ideal case, such 
studies require large amounts of simultaneous data in various EM bands;  unfortunately, this  is severely 
lacking for blazars. 

Thanks to the XMM-Newton satellite, which has instruments to observe simultaneously
a specific source in optical, UV, and X-ray bands, this limitation can be partially overcome. On searching  the 
complete archive of XMM-Newton,
we found that there were 20 occasions on which data in at least one optical-UV band as well as X-ray bands were taken
on same date for the BL Lac PKS 2155-304. These observations span a period of almost $12$ years.  They are thus ideal for studying the long-term flux and spectral variability of the source in the optical/UV/X-ray bands. We have  used these data sets to generate simultaneous broad-band SEDs for the low-energy hump and we have fitted these SEDs with  models to study the synchrotron emission mechanism and investigate how the various model parameters  vary with the source flux. 

This paper is structured as follows. In Section 2, we give a brief description of the XMM--Newton data 
reduction method. In Section 3 we discuss the long term variability in the light curves of different bands.  We describe our SED modeling in Section 4.  A discussion and our conclusions are given in Section 5.  

\section{The XMM-Newton Observations and data reduction}

Over the last $\sim$12 years,  the BL Lac PKS $2155-304$ has been observed by XMM-Newton on 
20 occasions. The journal of observations is given in Table 1. In the present work, we study  the data obtained from  the Optical Monitor (OM; Mason et al.\ 2001) and the European Photon Imaging Camera (EPIC) PN detector (Str{\"u}der et al.\ 2001). We did not consider the data from EPIC-MOS as the EPIC-PN data are more sensitive, and are less affected by photon 
pile-up effects. In all observations, the EPIC-PN detector  was operated in the  small window (SW) imaging mode. The OM  has three optical and three ultraviolet (UV) filters and can provide data in the optical/UV bands simultaneously with the X-ray observations.  In all, 20 observations with X-ray and at least one UV or optical band measurements are available in the archive.\\

We followed the standard procedure to reprocess the Observation Data File (ODF) with the XMM-Newton Science Analysis System (SAS) version 11.0.0 with the latest calibration files. We considered both single and double events 
{\it (PATTERN $\leq$ 4)} of good quality {\it (FLAG = 0)}. The source counts in each observation were accumulated from a circular region centered on the source and with a radius of 33$^{\prime \prime}$ to 40$^{\prime \prime}$. These radii have been chosen to sample most of the PSF according to the observing mode. Background counts were accumulated from a circular region of radius 45$^{\prime \prime}$ on the CCD chip where the source was located and was the least affected from  the source counts. The EPIC-PN redistribution matrix and effective areas were calculated with the {\tt rmfgen} and {\tt arfgen} tasks, respectively.

We checked for the high soft proton background periods which are caused by solar activity by generating a hard-band background light curve in the energy range 10$-$12 keV. We then defined as the ``good time interval" (GTI) those times where the hard band count rate was less than 0.4 ct/sec. We also investigated the possibility for photon pile-up effects, which may be  strong for a bright source such as PKS $2155-304$. To this end, we used the {\tt epatplot} SAS task. We found that nine observations were affected by photon pile-up. For these observations we excluded a circular region with a radius of 10$^{\prime \prime}$ centered on the source and we extracted the source counts in an annulus region which has an outer radius lying in the range of 33$^{\prime \prime}$ to 40$^{\prime \prime}$, depending on the position of the source on the chip. 

We used the {\tt efluxer} task to produce background subtracted, flux-calibrated EPIC-PN  X-ray spectra in physical units of erg cm$^{-2}$ s$^{-1}$. These spectra can be used to study the shape of the continuum X-ray emission in a model-independent way. On the other hand, their effective spectral resolution is degraded with respect to the intrinsic spectral resolution of EPIC-PN, but this is not a serious drawback in our case, as there are no narrow spectral features in the X-ray spectrum of this source. The final spectra are obtained between 0.3 -- 10 keV  with the default energy bins of the {\it efluxer} command. The Galactic hydrogen column density in the direction of PKS  $2155-304$ is $N_{\rm H} = 1.71\times 10^{20}$ cm$^{-2}$ which  has been calculated through the $N_{\rm H}$ calculator tool available online\footnote{http://heasarc.gsfc.nasa.gov/cgi-bin/Tools/w3nh/w3nh.pl};  this was developed by Lorella Angelini at the $HEASARC$ and uses the Dickey \& Lockman  (1990) density map. Although this is a rather low value, the X-ray spectra are expected to be significantly affected at energies below 0.5 keV. For that reason, we consider below only the flux-calibrated spectra at energies higher than 0.6 keV. We have used the 0.6-10.0 keV  X-ray flux in SED fittings.

The OM data were taken in the standard imaging mode. We reduced the OM data with the standard SAS routine {\tt omichain}. This routine provides  a {\it combolist} file with the source count rate and instrumental magnitudes for all the sources which are present in the filed of view. The PKS $2155-304$  fluxes corresponding to  six optical/UV filters were corrected for galactic reddening (E$_{B - V} =$ 0.019; Schlafly \& Finkbeiner 2011) with the standard reddening correction curve by Cardelli et al.\ (1989) and applied using equation (2) in Roming et al.\ (2009).  

\begin{figure*}
 \epsfig{figure=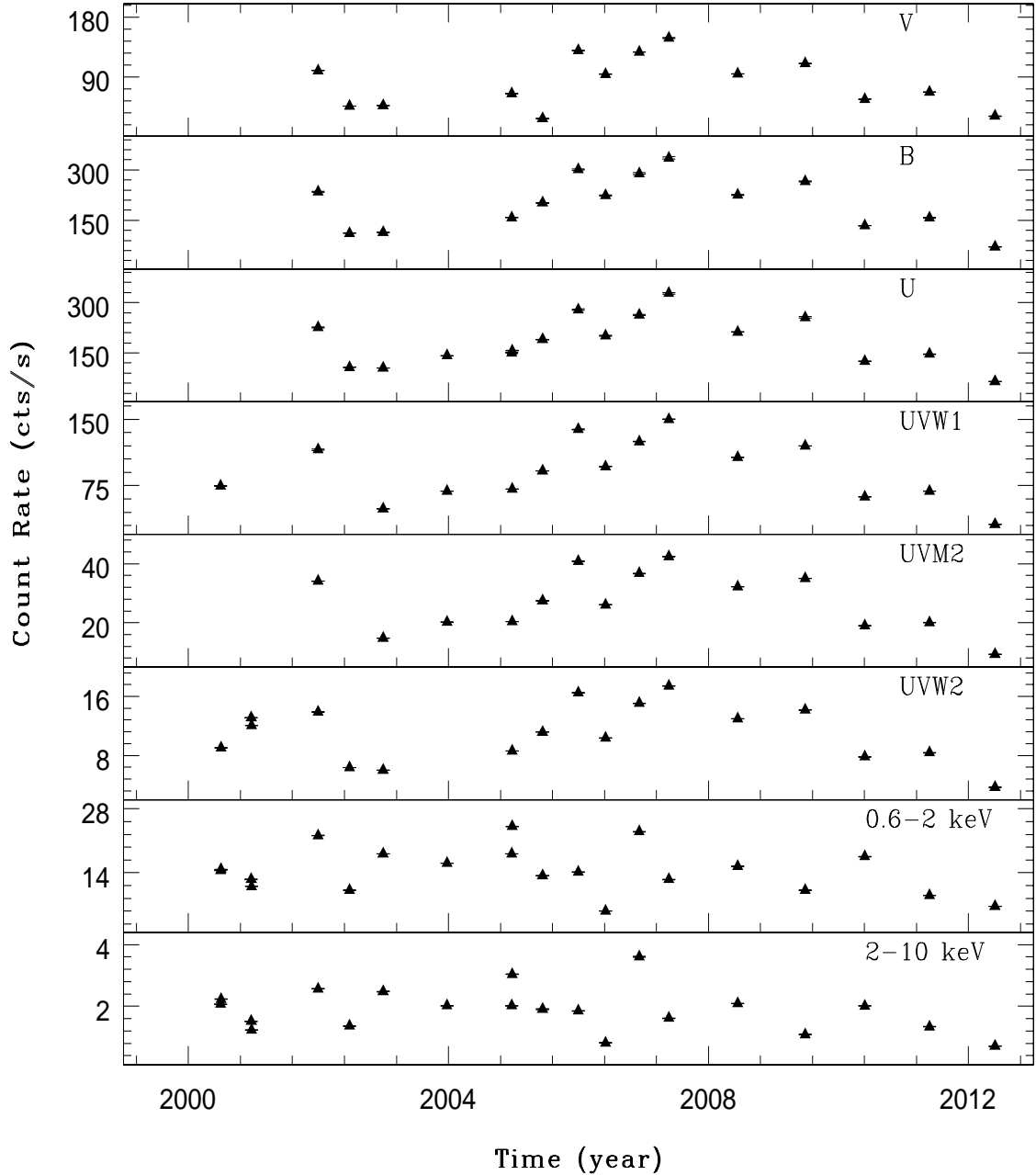,height=18.cm,width=16.cm,angle=0}
\caption{\scriptsize{Long term variability light curves for the XMM--Newton optical/UV and X-ray bands}}  
\label{fig:fig1}
 \end{figure*}

\section{The observed long-term light curves}

Fig.\ 1 shows the long term optical/UV/X-ray light curves of PKS $2155-304$, using the observations we studied in this work. 
The points in this figure indicate the average count rate of each observation in the various OM filters, in the 0.6--2 keV (``soft") and 2--10 keV (``hard") band. Obviously, the source is highly variable in all bands, over the time period of $\sim$12 years that the XMM--Newton observations were performed. We have estimated the rms variability amplitude (i.e. $\sqrt{\sigma^{2}/m^{2}}$ where $\sigma^2$ and $m$ are the variance, corrected for the experimental contribution, and mean of the light curve, respectively) for each light curve. The variability amplitude increased slightly going from the soft to hard X-ray bands but it decreased going from optical to UV bands. The values of rms variability amplitudes corresponding to the hard X-ray, soft X-ray, UVW2, UVM2, UVW1, U, B and V bands are $0.38$, $0.35$, $0.36$, $0.38$, $0.38$, $0.39$, $0.40$ and $0.47$, respectively. 

On visual inspection, the observed variations in the optical bands are well correlated with the variations in the UV bands. The same appears to be true with the variations detected in the soft and hard X-ray bands.  However, this is not the case when we compare the variability detected in the optical/UV bands and in the X-rays. Fig.\ 2 shows the UVW1 count rate plotted as functions of the 2--10 and 0.6--2 keV band measurements (upper and lower panels, respectively). Clearly, the flux variations in the UV and X-ray bands are not well correlated. \\

\begin{figure}
 \epsfig{figure=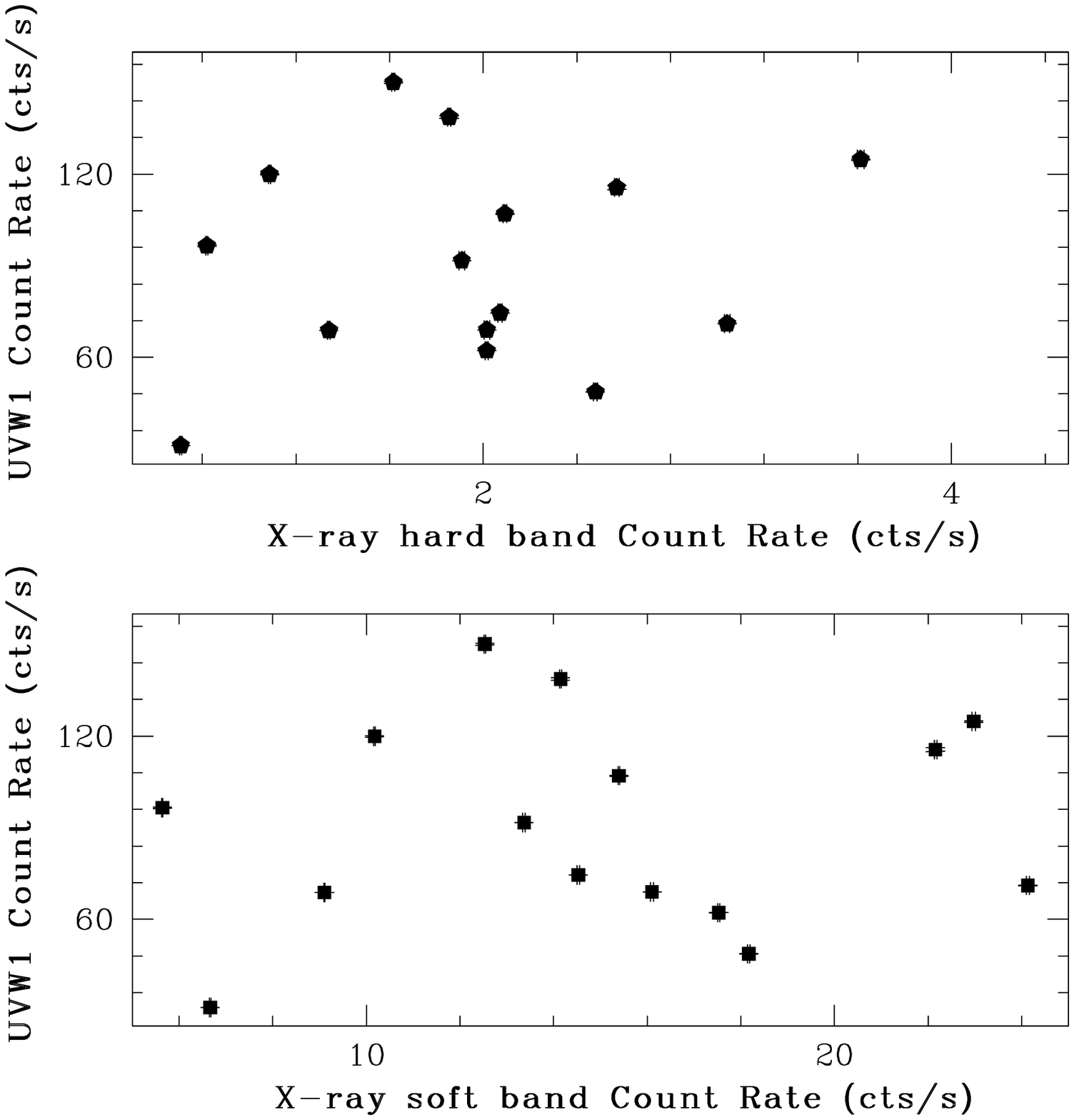,height=10.cm,width=8.cm,angle=0}
\caption{\scriptsize{UV versus soft (0.6-2.0 keV) and hard (2.0-10.0 keV) X-ray count rates}}
\label{fig:fig2}
 \end{figure}

\section{SED Modeling}

Fig.\ 3 shows three optical/UV to X-ray SEDs of PKS $2155-304$, using the mean optical and UV flux measurements in each  XMM-Newton observation and the flux calibrated EPIC-PN data we described in the previous section. They are representative of all the observed SEDs. The spectra cover a frequency range of over three orders of magnitude.  More importantly, as we have stressed earlier, the data in the optical and UV bands are simultaneous with those in the X-ray bands. Given the shape of the SED in the optical/UV and X-ray bands, the low energy synchrotron peak of this source is located between the energy bands sampled by the XMM-Newton OM and EPIC-pn observations. 

We fitted all SEDs with two models. The first one was a log-parabolic model. We also considered the case of a spectral 
model that has a power law shape at low energies, and then acquires a log-parabola form at higher energies, following 
Massaro et  al (2006) and Tramacere et al.\ (2009). We describe below the best-fit results in both cases.

\subsection{\bf Log-parabolic fits}
Log-parabolic models are parametrized with functions of the form $F(E)=KE^{-(\Gamma+bLog(E))}$, where $F(E)$ is the source flux in units of photons cm$^{-2}$ s$^{-1}$ keV$^{-1}$ at energy $E$ (see e.g. Massaro et al. 2004). The $\Gamma$ parameter is the photon index at 1 keV, and $b$ is a parameter that measures the spectral curvature. Since in our case the data are directly in flux density units (i.e. ergs cm$^{-2}$ s$^{-1}$), we decided to fit them with a model of the form: 

\begin{equation}
S(E)=K_S E^{-b[ \log(E/E_p) ]^2}, 
\end{equation}
where $S(E)=E^2F(E)$, and $K_S=E_p^2F(E_p)$. The model has three free parameters: $E_p$, which is the enery where the peak power is emitted (in units of keV),  $K_S$ (the model normalization), which indicates the power at $E_p$,  and the spectral curvature parameter  $b$.

Log-parabolic models best fit curved spectra which decrease symmetrically around their peak frequency and $b$  determines the curvature of the model around $E_p$. Similar models have been applied for a long time to parameterize blazar spectra in various energy bands. For example, Landau et al.\ (1986)  analyzed the SEDs of a sample of blazars in millimeter to UV bands and  found that the synchrotron emission of BL Lac sources were well fitted by a log-parabolic model. Krennrich et al.\ (1999)  also used the log-parabolic model to describe the spectral curvature of Mkn 421 in the TeV band,  while Giommi et al.\ (2002) applied it to the X-ray SEDs of 157 blazars observed by BeppoSAX.

\begin{figure*}
 \epsfig{figure=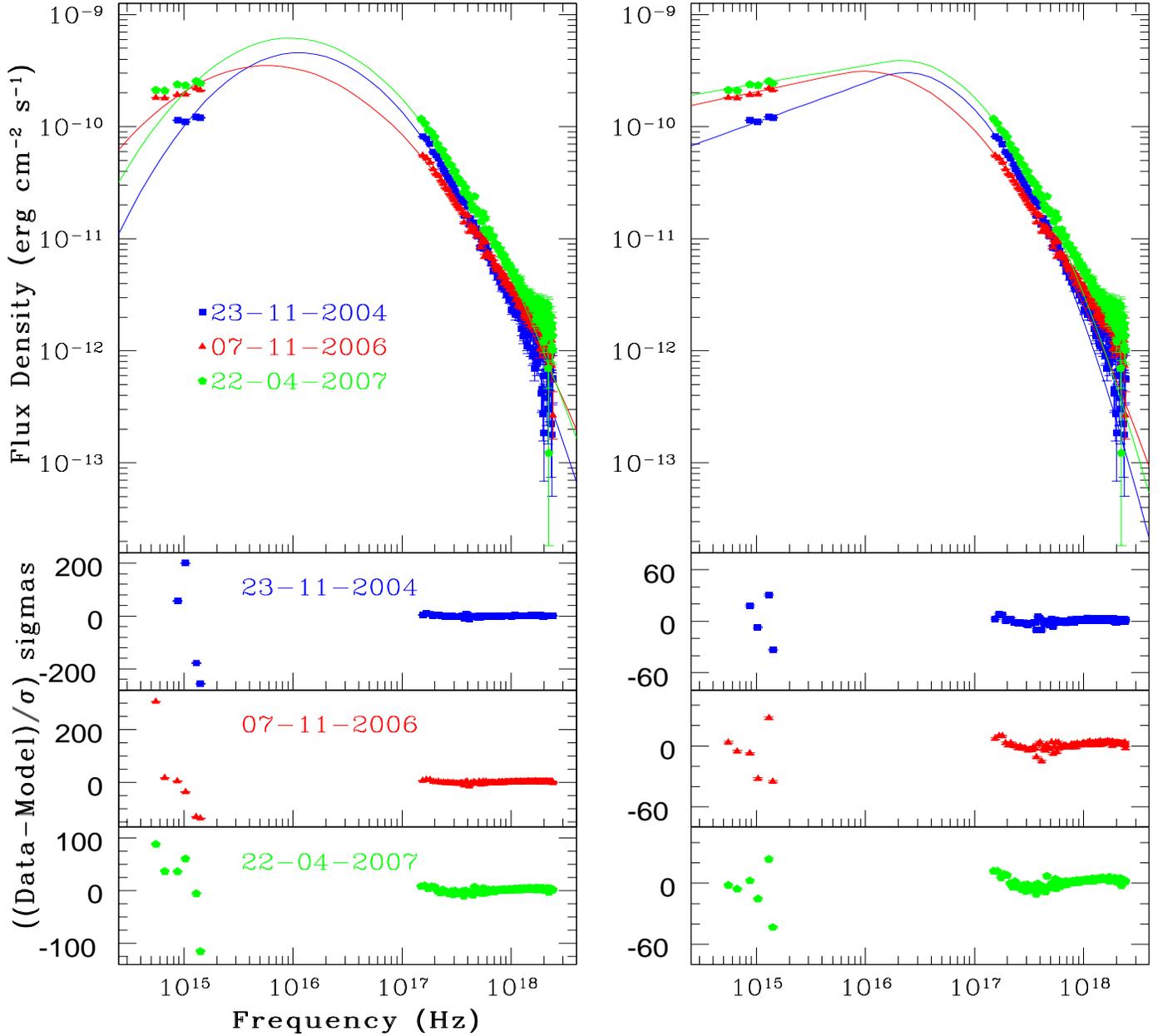,height=18.cm,width=20.cm,angle=0}
 \caption{\scriptsize{Example SEDs and best-fitting model curves for three observations. The best-fitting curves in the case of the log-parabolic model (together with the corresponding residuals) are  plotted in the upper and lower left panels, respectively. In the right hand panels, we plot the best-fitting curves (and the residuals) in the case of the power-law+log-parabolic model.}}
\label{fig:fig3}
 \end{figure*}

\begin{table}
\begin{center}
{\bf Table 2.}  The SED best-fit model parameter values for the log-parabolic model.  \\
\noindent 
\begin{tabular}{llcccrr}  \hline 
Observation ID & $K_S$ & $b$ & $E_{p} $ & $\chi^{2}$/dof \\
\hline
0124930101  &	 4.4	 & 0.50	 & 5.0    &  3257/122       \\
0124930201  &	 3.6	 & 0.49	 & 5.4    &  2489/122        \\
0080940101  &	 4.8	 & 0.53	 & 4.4    &  3094/122         \\
0080940301  &	 4.3	 & 0.54	 & 4.3    &  1423/122          \\
0124930301  &	 8.4	 & 0.60	 & 5.5    & 85765/127           \\
0124930501  &	 3.8	 & 0.49	 & 4.0    &  7850/124            \\
0124930601  &	 3.1	 & 0.55	 & 4.5    & 213989/127             \\
0158960101  &	 3.3	 & 0.55	 & 3.8    & 24141/123              \\
0158960901  &	 5.1	 & 0.56	 & 3.3    & 15388/124               \\
0158961001  &	 4.6	 & 0.59	 & 4.7    & 141156/125                \\
0158961101  &	 4.8	 & 0.50	 & 4.6    & 42881/125                 \\
0158961301  &	 6.1	 & 0.50	 & 3.8    & 276580/127                  \\
0158961401  &	 2.8	 & 0.43	 & 2.6    & 120717/127                  \\
0411780101  &	 3.5	 & 0.40	 & 2.4    & 132211/127                  \\
0411780201  &	 6.2	 & 0.52	 & 3.9    & 28801/127                  \\
0411780301  &	 6.0	 & 0.52	 & 4.7    & 187076/127                  \\
0411780401  &	 6.2	 & 0.57	 & 4.0    & 288613/127                  \\
0411780501  &	 3.0	 & 0.54	 & 3.8    & 157672/127                  \\
0411780601  &	 3.1	 & 0.47	 & 4.1    & 175309/127                  \\ 
0411780701  &	 1.5	 & 0.55	 & 3.3    & 137475/126                  \\
   
\hline
\end{tabular}
\end{center}

$K_S$:  normalization constant (peak power) in units of $10^{-10}$ erg cm$^{-2}$ s$^{-1}$. \\
$b$: spectral curvature parameter. \\
$E_{p}$:  energy at which the peak power is emitted in units of $10^{-2}$ keV  \\ 
  
\end{table} 

The best model fit results are listed in Table 2, together with the best-fitting $\chi^2$ values (and the degrees of freedom -- dof). Obviously, the best-model fits are not statistically acceptable for any of the 20 SEDs. (It is for this reason that we do not provide errors on the best-fit model parameter values.) Examples of the quality of the  model fits are shown in Fig.\  3. The solid lines in the upper left hand panel of Fig.\ 3 indicate the best-fitting log-parabolic models to the SEDs that are plotted in the same panel, and in the left lower panels of the same Figure we also plot the best-fit residuals.

Fig.\ 3 indicates that there are systematic discrepancies between the best-fitting models and the data at 
both high ($>10^{18}$ Hz) and low ($<10^{15}$ Hz) frequencies. 
These discrepancies could be the  major reasons  for the large $\chi^2$ values. In the high frequency end, the observed SED is always flatter than the best-fit models. This spectral flattening could be caused by the fact that the IC component starts contributing to the emission observed above $\sim 10^{18} Hz$ (i.e. $\sim 4-5$ keV). In the low frequency end, the observed flux is higher than the model flux in almost all cases (see for example the 22/04/2007 SED in Fig.\ 3). 

There are three obvious physical possibilities for the discrepancy at lower frequencies. The first would be the contribution of the host galaxy emission, which should be more significant in the optical band. However, if this contamination were to be important the discrepancy should be much smaller at high flux states since we can safely assume the host galaxy emission is constant; this is not the case. The second possible physical explanation would be that the emission from the broad line region (BLR) and/or the underlying accretion disk is variable, and contributes significantly in the low frequency part of the observed SED. 

A third possibility is that a log-parabolic model is not actually the true underlying physical model for the broad band, optical/UV up to X-ray SED of the source. For that reason, we also investigated the possibility that the low energy segment of the PKS $2155-304$ UV to X-ray spectra follows a single power law and the log-parabolic bending becomes apparent only above a ``critical", turn-over energy, $E_c$.

\subsection{Power-law plus log-parabolic (PLLP) fits}   
In the case of  spectra in units of photons cm$^{-2}$ sec$^{-1}$ keV$^{-1}$) this model is defined as: 
$F(E)=K (E/E_c)^{-\Gamma}$,  at energies below $E_{\rm c}$, and 
$F(E)=K (E/E_{\rm c})^{[-\Gamma-b \log(E/E_{\rm c})]}$ at energies higher than $E_{\rm c}$ ($\Gamma$ is the photon index). For spectra in flux density units (like our case), the above equations become: 

\begin{equation}
S(E)=K_{\rm S} (E/E_{\rm c})^{-\alpha'}, E\leq E_{\rm c},
\end{equation}

\begin{equation}
S(E)=K_{\rm S}  (E/E_c)^{[-\alpha'-b \log(E/E_c)]}, E>E_c,
\end{equation}

where $K_{\rm S} $ is the model normalization ($K_{\rm S} =KE_{\rm c}^2$), and $\alpha'=\Gamma-2$ is the spectral index of the SED in flux density units (i.e., $S(E)=E^2F(E)$; we chose to denote the spectral index with $\alpha '$ in order to distinguish it from the usual spectral slope, $\alpha=\Gamma-1$, which applies to SEDs in power over keV units). In this model, the energy where the peak power, $S_{\rm p}$, is emitted is given by: $E_{\rm p}=E_{\rm c} 10^{-\alpha'/2b}$. 

\begin{table*}
\begin{center}
{\bf Table 3.}  The SED best-fitting model parameter values for the power-law+log-parabolic model. (Numbers after the slash indicate the best fitting results in the case of the model fits to the SED data up to only $5\times 10^{17}$Hz.)  \\
\noindent 
\begin{tabular}{lccccc}  \hline 
Observation ID & $K_{\rm S}$ & $\nu_{\rm c}$ & $\alpha'$ & $b$ & $\chi^{2}$/dof \\
\hline
0124030101 & 2.2/3.1($\pm 0.3$)  & 2.3/5.0($\pm 0.9$)   &  -0.82/-0.64($\pm 0.13$)     &  0.54/0.62($\pm 0.03$)   &  1208/122  \&  394/25   \\ 
0124930201 & 1.7/2.0($\pm 0.2$)  & 2.0/2.3($\pm 0.5$)   &  -0.86/-0.92($\pm 0.12$)     &  0.52/0.57($\pm 0.03$)   &  1847/122  \&  646/25  \\ 
0080940101 & 2.8/3.2($\pm 0.3$)  & 2.4/3.0($\pm 0.5$)   &  -0.77/-0.76($\pm 0.14$)     &  0.56/0.62($\pm 0.03$)   &  1171/122  \&  379/25 \\   
0080940301 & 2.5/3.0($\pm 0.3$)  & 2.3/2.9($\pm 0.6$)   &  -0.81/-0.79($\pm 0.12$)     &  0.58/0.64($\pm 0.03$)   &  1012/122  \&  245/25   \\ 
0124030301 & 4.8/5.2($\pm 0.5$)  & 11.6/14.2($\pm 2.8$) &  -0.40/-0.39($\pm 0.08$)   &  0.70/0.80($\pm 0.04$)   &  5294/127  \&  4058/30  \\ 
0124930501 & 2.1/2.2($\pm 0.3$)  & 10.0/16.0($\pm 3.6$)  &  -0.40/-0.35($\pm 0.07$)    &  0.63/0.80($\pm 0.04$)   &  1000/124  \&  280/27   \\ 
0124030601 & 1.5/1.6($\pm 0.3$) & 16.6/20.6($\pm 3.8$) &  -0.24/-0.23($\pm 0.08$)   &  0.71/0.83($\pm 0.04$)   &  8266/127  \&  7030/30  \\
0158960101 & 2.3/2.6($\pm 0.3$)  & 4.2/5.3($\pm 1.6$)   &  -0.53/-0.53($\pm 0.11$)     &  0.60/0.67($\pm 0.03$)   &  1029/123  \&  412/27  \\
0158960901 & 2.6/2.7($\pm 0.3$)  & 8.4/11.2($\pm 2.2$)  &  -0.38/-0.35($\pm 0.06$)    &  0.69/0.79($\pm 0.04$)   &  919 /124  \&  335/27  \\
0158961001 & 2.6/2.8($\pm 0.2$)  & 11.8/14.4($\pm 3.2$) &  -0.35/-0.35($\pm 0.08$)   &  0.73/0.83($\pm 0.04$)   &  3310/125  \&  2726/28  \\ 
0158961101 & 3.2/3.6($\pm 0.4$) & 5.1/6.8($\pm 1.3$)   &  -0.49/-0.49($\pm 0.10$)     &  0.55/0.64($\pm 0.03$)   &  5288/125  \&  4686/29  \\ 
0158961301 & 4.1/4.4($\pm 0.4$)  & 7.6/9.8($\pm 1.5$)   &  -0.30/-0.30($\pm 0.09$)     &  0.57/0.66($\pm 0.03$)   &  6965/127  \&  5727/30  \\
0158961401 & 2.0/2.1($\pm 0.3$)  & 6.6/9.3($\pm 1.9$)   &  -0.15/-0.15($\pm 0.08$)     &  0.48/0.58($\pm 0.03$)   &  4726/127  \&  2782/30 \\ 
0411780101 & 2.8/3.0($\pm 0.3$)  & 4.4/6.1($\pm 1.2$)   &  -0.21/-0.21($\pm 0.10$)     &  0.43/0.52($\pm 0.03$)   &  4533/127  \&  3473/30 \\
0411780201 & 3.8/3.8($\pm 0.5$)  & 11.8/16.8($\pm 3.0$) &  -0.18/-0.16($\pm 0.11$)   &  0.61/0.75($\pm 0.03$)   &  4792/127  \&  3100/31 \\ 
0411780301 & 3.8/4.2($\pm 0.4$)  & 7.8/10.0($\pm 1.9$)   &  -0.39/-0.39($\pm 0.10$)     &  0.59/0.68($\pm 0.04$)   &  6042/127  \&  4577/30 \\ 
0411780401 & 3.8/4.0($\pm 0.4$)  & 8.9/10.8($\pm 1.8$)  &  -0.32/-0.32($\pm 0.09$)    &  0.66/0.74($\pm 0.04$)   &  5000/127  \&  3980/31 \\
0411780501 & 1.9/2.1($\pm 0.2$)  & 6.4/7.7($\pm 1.8$)   &  -0.38/-0.38($\pm 0.10$)     &  0.60/0.67($\pm 0.03$)   &  6452/127  \&  5299/31 \\
0411780601 & 2.0/2.0($\pm 0.3$)  & 10.0/12.7($\pm 2.5$) &  -0.26/-0.25($\pm 0.09$)   &  0.55/0.64($\pm 0.03$)   &  8006/127  \&  6354/31 \\
0411780701 & 0.8/0.9($\pm 0.2$)  & 9.9/12.5($\pm 2.0$)  &  -0.23/-0.23 ($\pm 0.10$)   &  0.66/0.77($\pm 0.04$)   &  3721/126  \&  3101/30 \\

\hline
\end{tabular}
\end{center}

$K_{\rm S}$:  normalization constant in units of $10^{-10}$ erg cm$^{-2}$ s$^{-1}$. \\
$\nu_c$:  turn-over frequency in units of $10^{15}$ Hz. \\ 
  
\end{table*} 

The best model fitting results are listed in Table 3,  together with the best-fitting $\chi^2$ values and dof. Although the $\chi^2$ values have decreased significantly when compared to the $\chi^2$ values in the case of the log-parabolic best model fits, these models are still not statistically acceptable for any of the 20 SED.  Examples of the quality of the model fits are also shown in Fig.\  3. The solid lines in the upper right hand panel in this figure indicate the best-fit PLLP models to the SEDs that are plotted in the same panel, and in the right lower panels of  the same Figure we plot the respective best-fit residuals.

This time, the best-model fits do not over-predict the UV spectra at low frequencies, but the X-ray band residuals display a ``wavy" pattern. This is the result of the fact that the model does not fit well the X-ray data above $\sim 5\times 10^{17}$ Hz. The $\chi^2$ values decrease even more, although the reduced $\chi^2$ values increase significantly in some cases. A major reason for this is the extremely small error bars in the optical/UV part of the spectra. Indeed, the discrepancy between the best-fit model and the observed SEDs (i.e. the ratio “(data-model)/data”) is of the order of  $\sim$ 5\% in the UV band, and even smaller in the X-ray band (except from the $3.5-4\times 10^{17}$ Hz region, where we observe discrepancies of the order of $10-20$ \% in all spectra), even in the case of the SEDs with the highest $\chi^2$ values. In fact, if we assume that the error of the SED points in all cases is equal to 5\% of the SED values

These reduced $\chi^2$ values improve even more but are still not statistically acceptable. A major reason for this is the extremely small error bars (especially in the optical/UV part of the spectra). Indeed, the discrepancy between the best-fit model and the observed SEDs (i.e. the ratio ``data/model") is of the order of $\sim 5$\% in the UV band, and even smaller in the X-ray band (except from the $3.5-4\times 10^{17}$ Hz region, where we observe discrepancies of the order of $10-20$ \% in all spectra). In fact, if we assume that the error of the SED points in all cases is equal to 5\% of the SED values, then the best-fit models to all spectra are now acceptable (with reduced $\chi^2$ values of the order of $1-2$). In this case, we can also estimate the 1-$\sigma$ error for the best-fit parameter values. These errors are indicated in {\bf Table 3} by the numbers in the parentheses next to the best-fit results in the case of the SED fits up to the $5\times 10^{17}$ Hz.

\begin{figure*}
 \epsfig{figure=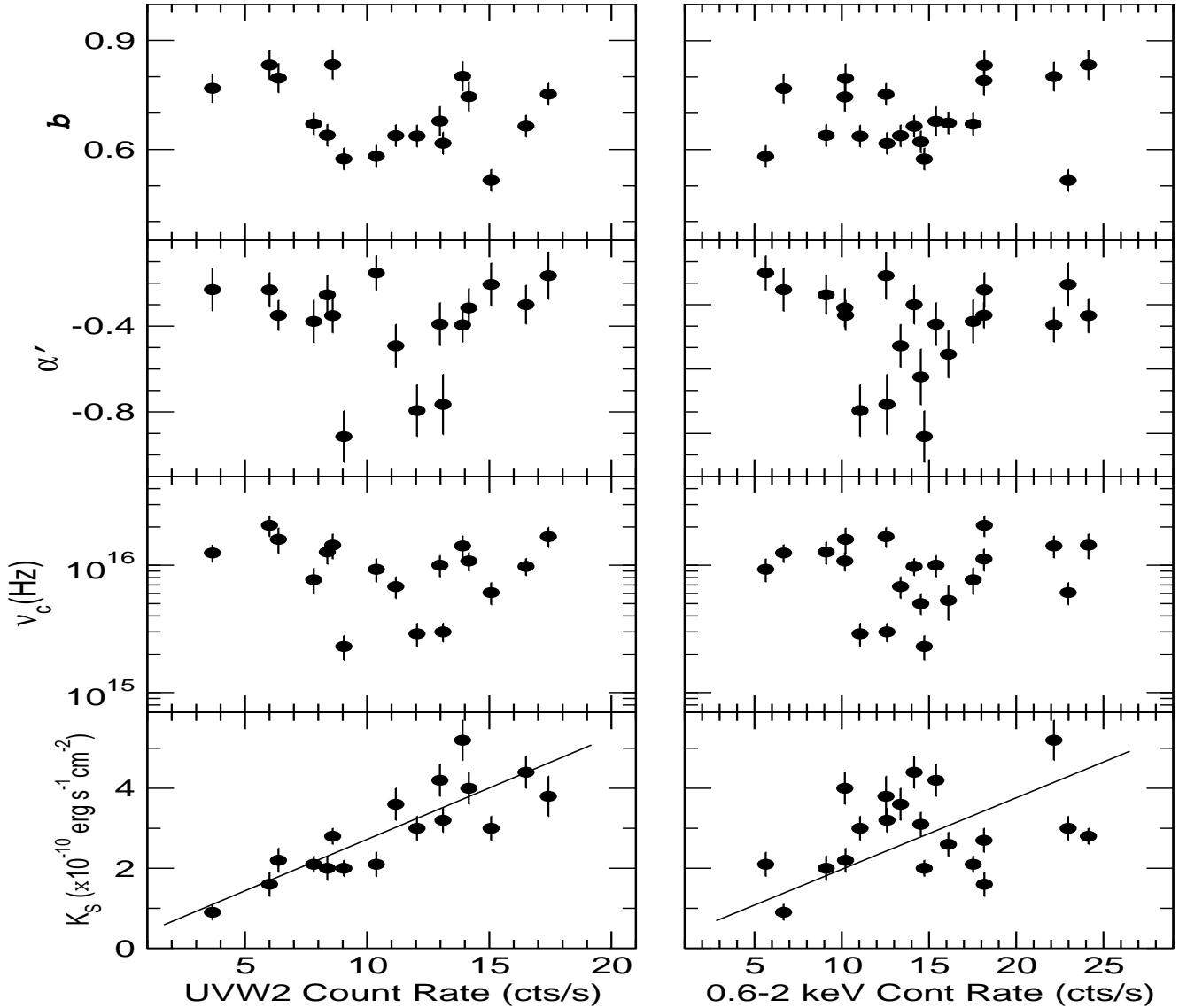,height=18.cm,width=20.cm,angle=0}
 \caption{\scriptsize{Plots of the best-fit values of the spectral curvature $b$, spectral slope $\alpha '$, turn-over frequency $\nu_{\rm c}$ and normalization constant $K_{\rm S}$, as function of the UVW2 and the 0.6--2 keV count rates (left and right panels, respectively). The solid line in the lower panels indicates lines with a slope of one, and are plotted for illustration purposes (see text for details).}}
\label{fig:fig4}
 \end{figure*}

We used the PLLP best-fit results to investigate correlations between the best-fit model parameters and the observed UV and soft X-ray count rates. Fig.\ 4 shows plots of the best-fit model parameters with the observed UVW1 and 0.6--2 keV count  rates (left and right panels, respectively). To quantify the correlation we used both the frequently used Pearson's $r$ as well as the non-parametric Kendall's $\tau$.  Only the model normalization (i.e., $K_{\rm S}$) is positively correlated with the UV flux (see lower left panel in Fig.\ 4). The correlation is statistically significant, and rather strong: Pearson's $r$ is 0.84 ($\tau=0.65$)  and the probability that this value appears by chance is $P_{\rm null}=2.5\times10^{-5}$ ($3\times 10^{-4}$).  In order to emphasize even more the correlation between the model normalization and the UV flux, the solid line in the bottom left panel of Fig.\ 4 indicates a straight line with a slope of unity. This is not the best fit to the data plotted in the same panel, but we plot it there in order to indicate that such a line appears to describe well the relation between the data plotted. Therefore, the UV flux variations could indeed be, to a large extent, proportional to the model normalization variations; i.e.. to a first approximation, the UV flux is simply responding to the model normalization variations, without being affected by the other model parameter changes.

On the other hand, $K_{\rm S}$  does not correlate significantly with the soft X-ray flux (and neither does any of the other best-fit model parameter values). In the lower-right panel of Fig.\ 4 we also plot the one-to-one line. There may exist a rough positive correlation between the two quantities, but both Pearson's $r$ and Kendall's $\tau$ imply that this correlation is not statistically significant. This result explains the lack of correlation we observe between the UV and X-ray fluxes (see Fig.\ 1). While the UV flux responds mainly to the model normalization, the X-ray flux must also be significantly affected by spectral shape variations as well (i.e. variations of $b$ and $E_c$).

We also investigated the correlations within the model parameters. The only significant correlations that we found are those between $\nu_{\rm c} $ (the turn-over frequency which corresponds to $E_{\rm c}$) and $\alpha'$ (Pearson's $r=0.73$, Kendall's $\tau=0.5$, and P$_{\rm null}=2.2\times 10^{-4}$ and $2.3\times 10^{-3}$, respectively) and between $\nu_{\rm c}$ and $b$ ($r=0.80, \tau=0.65,$ P$_{\rm null}=2.7\times 10^{-5}$, and $5.7\times 10^{-5}$, respectively). Fig.\ 5 shows a plot of $\alpha '$ and $b$ versus $\nu_{\rm c}$. Our results indicate that  flatter and more curved spectra are associated with higher turn-over frequencies. 

The upper panel in Fig.\ 6 shows a plot of the model curvature $b$ as a function of $\nu_{\rm p}$ (the frequency at which the maximum power is emitted). Not surprisingly, given the correlation between $\nu_{\rm c}$, $\alpha '$, and $b$, the parameters $b$ and $\nu_{\rm p}$ are also strongly correlated. The correlation is positive, in the sense that as the peak power increases, the spectral curvature increases as well. The dashed line in the same panel indicates the best-fit line to the data (in the log--log space). The fit has been performed using the ``ordinary least-squares bisector" method of Isobe et al. (1990). The best-fit result indicates that:  $b \propto \nu_{\rm c}^{0.48\pm 0.04}$.  The error on the best-fit slope value indicate that the positive correlation between the parameters (in the log--log space) is significant at a level much higher than $3\sigma$. 

The lower panel in the same Figure shows $S_{\rm p}$ (i.e. the maximum emitted power) plotted as a function of $\nu_{\rm p}$.  The two parameters appear to be loosely 
anti-correlated ($r=-0.21, \tau=-20$, but $P_{\rm null}\sim$0.2--0.3, in both cases). However, when we fit the data (in log-log space, we find that: $S_{\rm p}\propto \nu_{\rm c}^{-1.23\pm0.23)}$. This result indicates that, if there is a correlation between these two parameters, it is an  anti-correlation: as the  maximum power emitted increases, the frequency at which it is emitted decreases. 

\begin{figure}
\includegraphics[trim=0.0cm 0.0cm 0.0cm 0.0cm, clip=true, scale=0.5]{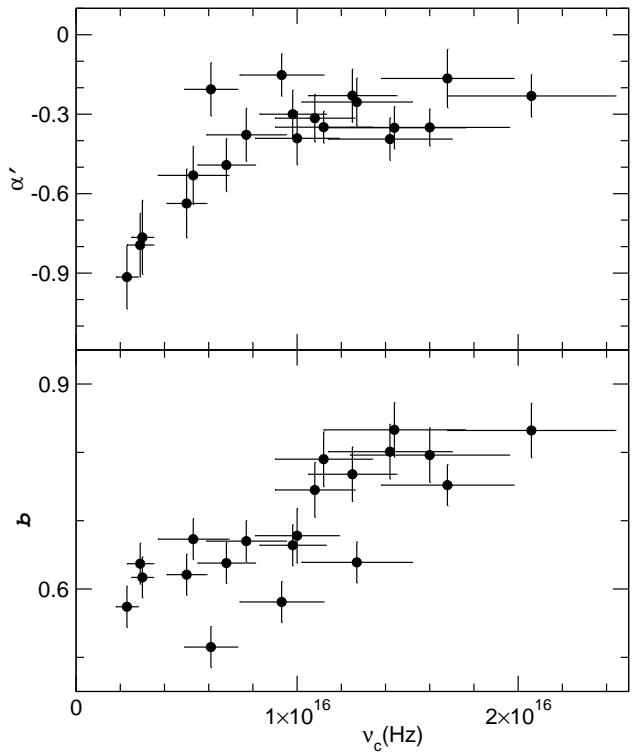}
\caption{\scriptsize{The best fitting PLLP  $\alpha '$ and $b$ values plotted as a function of the turn-over frequency, $\nu_{\rm c}$.}}
\label{fig:fig5}
\end{figure}

\begin{figure}
\includegraphics[trim=0.5cm 6.5cm 1.0cm 3.1cm, clip=true, scale=0.5]{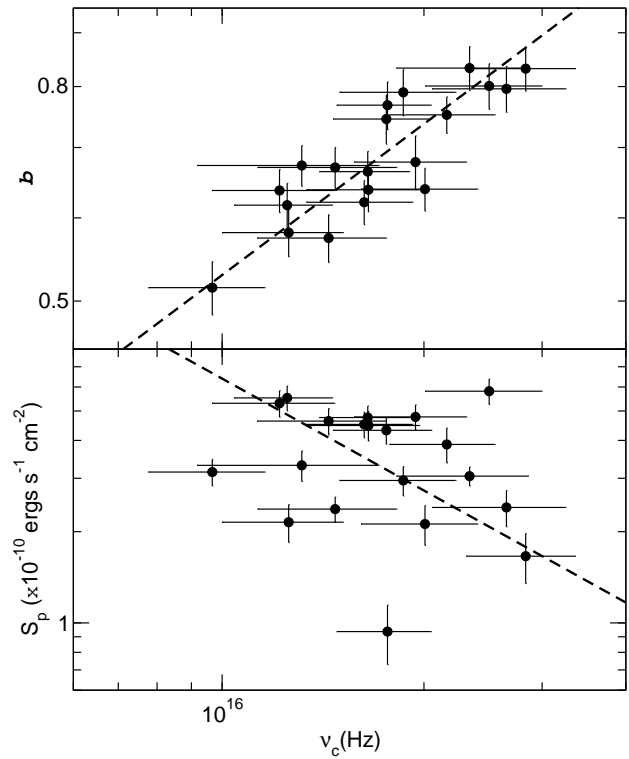}
\caption{\scriptsize{The $b$ and $S_{\rm p}$ parameter values plotted as a function of the turn-over frequency, $\nu_{\rm c}$.}}
\label{fig:fig6}
\end{figure}

\section{Discussion and Conclusions}

We have studied 20 archival XMM-Newton observations of PKS $2155-304$ which have been performed in a period of over twelve years from 2000 to 2012. These observations can be useful  in the study of the long-term optical/UV and X-ray variability of the source, not just because their number is large, but also because they allow us to study the flux and spectral variability of the source, over a  broad frequency range, {\it simultaneously}.  Our main results can be summarized as follows: 

(1) The source is variable at all bands on time scales of years. The amplitude of the rms variability is of the order of $\sim 35-45$\% at all bands. We did not observe any extreme activity taking place during these observations. The variability amplitude slightly increases from the soft to the hard X-ray band, and  decreases from the optical to the UV bands. 

(2) The optical/UV band fluxes increase and decrease in phase, i.e. the optical/UV band variations are well correlated. However, the X-ray and optical/UV fluxes are not correlated. 

We then used (i) a log-parabolic (LP) model and (ii) a power-law plus log-parabolic (PLLP) model to fit the broad band SEDs. The LP model fits are not formally acceptable. Massaro et al.\ (2004) and Tramacere et al.\ (2009) have found that a LP model can well fit the optical/UV and X-ray spectra individually, but could not  fit  the combined optical/UV and X-ray bands.  Our results are in agreement with their results.

The fits improve in the case of the  PLLP model, but there are still significant discrepancies above $\sim 4$ keV in the X-rays, most probably due to the increased contribution of the IC component at larger energies. We repeated the fits using data up to $5\times 10^{17}$ Hz only. These model fits appear to describe rather well the overall shape of the optical/UV/X-ray spectrum of the source at the $\sim 5$\% level, i.e. the data/model ratio is typically between 0.95 and 1.05 at all energies up to $5\times 10^{17}$ Hz). If we increase the error of the data points to this level, then the  PLLP model fits the data well. 

If we accept that a PLLP model can parametrize the optical/UV and X-ray SED of the source then our results from the SED fitting of all observations can be summarized as follows: 

(3) The turn-over frequency correlates {\it positively} with the model spectral slope, and with the curvature parameter, $b$:  as the turn-over energy increases, the spectrum steepens, and the curvature parameter increases. 

(4) Due to the above mentioned correlations, the peak frequency, $\nu_{\rm p}$ and the curvature parameter $b$ are also positively correlated. As $\nu_{\rm p}$ shifts to higher energies, the spectral curvature also increases following the relation: $b\propto \sqrt{\nu_{\rm p}}$.

(5) We do not observe a strong correlation between the peak power, $S_{\rm p}$, and the peak frequency, $\nu_{\rm p}$. If there is a relation between these two parameters, it is most probably an  anti-correlation, in the sense that as the peak luminosity decreases, the peak frequency shifts to to higher energies, roughly according to the relation:  $S_{\rm p} \propto 1/\nu_p$.

Massaro et al.\ (2008) considered a sample of blazars and observed an
anti-correlation between $E_p$ and $b$ for five TeV blazars including PKS $0548-322$, 1H $1426+428$, Mrk 501, and 1ES $1959+650$. They also found a positive correlation between the spectrum peak power, $S_p$,  and $E_p$. These correlations and anti-correlations were based on the results when a log-parabolic model was used to fit 
the X-ray spectra only. Clearly, our results regarding the relations between $E_p, b$ and $S_p$ are contrary to those reported by Massaro et al.\ (2008). However, they also studied PKS 2155-204, and found that the $S_p - \nu_p$ and $b-\nu_p$ relations in this object were different than the same relations for the other objects in their sample results. In fact, the observational relations presented in their Fig.\ 8 are quite similar to our plots shown in Fig.\ 6. 

From a phenomenological point of view, the PLLP model can be explained if the the electron distribution at low energies follows a power-law up to a turn-over energy, and a log-parabolic shape at higher energies. If that is the case, our results indicate that this low-energy power-law branch is always present in PKS 2155-304. Furthermore, we can constrain the typical slope of the power-law energy distribution of the electrons, $s$, using the the well-known relation between $s$ and $\alpha '$: $\alpha '=(s-3)/2$. For the mean spectral index value of $\alpha '\sim -0.41$ in our case, we estimate that $s\sim 2.2$. This slope is fully consistent  with predictions of models which assume first-order Fermi acceleration as being the primary acceleration mechanism in most collisionless magnetohydrodynamic (MHD) shocks, as has been shown both analytically (e.g.\ Bell 1978, Kirk et al.\ 2000) and numerically (e.g.\ Bednarz \& Ostrowski 1998; Baring et al. 1999; Ellison \& Double 2004).

The presence of the log-parabolic branch in the electron distribution can be explained as in Massaro et al.\ (2004). These authors have shown that when the acceleration efficiency of particles is inversely proportional 
to the energy itself, then the energy distribution approaches a log-parabolic shape. They proposed that  
the log-parabolic spectra are naturally produced when the statistical acceleration probability have an energy dependency. 
According to this model the curvature, $r$, is related to the fractional acceleration gain $\epsilon$ by 
$r \propto [\rm{log} \epsilon]^{-1}$ and $E_{\rm P} \propto \epsilon$, where $E_{\rm P}$ is the peak energy.
This produces a negative trend in the relation between $E_{\rm P}$ and $b$ (see Tramacere et al.\ 2009). 

An alternative explanation of the above trend is 
provided by the stochastic acceleration framework that includes a momentum diffusion term. The diffusion term plays a 
crucial role in the broadening of the spectral shape of the electrons (Kardashev et al.\ 1962; 
Massaro et al.\ 2006).  Tramacere et al.\ (2009) showed that the log-parabolic spectrum results from the evolution
of a mono-energetic or quasi-mono-energetic particles injection under a Fokker-Planck 
equation with a momentum-diffusion term. Kardashev 
et al.\ (1962) have shown that the curvature term $r$ is inversely proportional to the diffusion term $D$ and the time $t$:
$r \propto \frac{1}{Dt}$. 
This relation leads to the following connection between the peak frequency, the peak energy of the  electron distribution, 
$\gamma_{p}$, and the spectral curvature $b$ (Eqn.\ 5 of Tramacere et al.\ 2009): ${\rm ln} (E_{\rm P})=2 {\rm ln} (\gamma_{p})+3/(5b)$.

Hence both the fractional acceleration gain term, $\epsilon$, and the momentum diffusion term, $D$ predict an  
anti-correlation  between $E_{\rm P}$ and $b$. However, this opposite to what we observe. As we showed in Section 4, if there is a relation between $E_{\rm P}$ and $b$, this is a {\it positive}, and not a negative one.

The inability of the model to provide acceptable fits to the broad band optical/UV/X-ray SEDs of PKS $2155-304$, as well as the positive correlation between $E_{\rm P}$ and $b$ that we observe,  perhaps indicates that the optical/UV and X-ray emission in this source are produced by two different populations of leptons. Optical/UV emission may be produced by slow leptons and the X-ray emission may due to emission from much more energetic leptons, which may have been  accelerated through the energy dependent particle acceleration mechanism. This possibility could also explain the fact that the optical/UV bands are well correlated with each other but not correlated with X-ray bands.  
Another possibility for the positive correlation between $E_{\rm P}$ and $b$  can arise within the  stochastic acceleration framework if the cooling losses  successfully compete with the acceleration and diffusion components  (Tramacere et al.\ 2011).   However, only if we can analyze more SEDs, preferably including a wider range of EM bands,  might the correlations between these parameters  be clarified.   That would definitely increase our understanding of the emission processes that dominate the spectra of blazars.
 
\section*{Acknowledgments}
We are grateful to the anonymous referee for insightful suggestions. This work is based on the observations obtained with XMM-Newton, an ESA science mission with instruments and contributions directly funded by the ESA member states and NASA.

\end{document}